\documentclass[5p]{elsarticle}

\usepackage{graphicx,color}

\usepackage{dcolumn}
\usepackage{bm,ulem}
\usepackage{amsmath,amssymb}
\begin{document}
\begin{frontmatter}

\title{Tetragonal CuMnAs alloy: role of defects}

\author[label1]{F. M\'aca}\ead{maca@fzu.cz}
\author[label1]{J. Kudrnovsk\'y}
\author[label2,label3]{P. Bal\'a\v{z}}
\author[label1]{V. Drchal} \address[label1]{Institute of Physics ASCR, Na Slovance 2, CZ-182 21 Praha 8, Czech
Republic}
\author[label2]{K. Carva}
\author[label2]{I. Turek}
\address[label2]{Charles University, Faculty of Mathematics and Physics, Department
of Condensed Matter Physics, Ke Karlovu 5, CZ-121 16 Praha 2, Czech Republic} \date{\today}
\address[label3]{IT4Innovations Center, VSB Technical University of Ostrava, 17. listopadu 15, CZ-708 33 Ostrava-Poruba, Czech Republic}

\begin{abstract}

%Electronic, magnetic, and transport properties of
 The antiferromagnetic (AFM) CuMnAs alloy with tetragonal structure is a promising material
 for the AFM spintronics. The resistivity measurements indicate the presence of defects
 about whose types and concentrations is more speculated as known. We confirmed vacancies
 on Mn or Cu sublattices and Mn$_{\rm Cu}$ and Cu$_{\rm Mn}$ antisites as most probable defects in
 CuMnAs by our new ab initio total energy calculations.  We have estimated resistivities
 of possible defect types as well as resistivities of samples for which the X-ray structural
 analysis is available. In the latter case we have found that samples with Cu- and Mn-vacancies
 with low formation energies have also resistivities which agree well with the experiment.
 Finally, we have also calculated exchange interactions and estimated the N\'eel temperatures
 by using the Monte Carlo approach. A good agreement with experiment was obtained.

\end{abstract}

%\pacs{75.25.+z,75.30.Et,75.47.Np,75.50.Ee}

\end{frontmatter}

\section{Introduction}

The CuMnAs crystallizes in the orthorhombic phase \cite{cumnas-or}, the tetragonal phase can be prepared
only under special conditions by separation from an ingot or as a film by the molecular beam epitaxy (MBE)
 on a suitable substrate, namely GaAs(001) and GaP(001) \cite{cumnas-2}. The tetragonal antiferromagnetic (AFM) CuMnAs phase
has attracted recently large interest as promising material for
applications in the so-called AFM spintronics \cite{cumnas-2,afm-spin,cumnas-1,cumnas-3}.

 The first experiments \cite{cumnas-1,cumnas-str} found
basic characteristics including structural parameters that were used in first-principles
calculations assuming an ideal structure
without defects \cite{cumnas-1}. The experiment on real samples gave the
 residual resistivity around 90~$\mu\Omega$cm for low temperature $T\approx 5$~K and
 160~$\mu\Omega$cm for room temperatures \cite{cumnas-1}.
The measurements indicate local Mn moments around 3.6~$\mu_{\rm B}$ at room temperature
\cite{cumnas-1}, the N\'eel temperature around 480~K has been found \cite{cumnas-tn}.

  The residual resistivity indicates the presence
of defects whose origin and concentrations are known only very approximately \cite{cumnas-str}.
Our previous study \cite{our-PRB} identified the most probable defects but the evaluation of their formation energies
 represents a challenge for the theory. The same concerns also an estimate of the residual
resistivity and the N\'eel temperature.

In this contribution we will present a more accurate estimate of formation energies and discuss in detail resistivities
of samples for which possible defect structures were obtained by the X-ray analysis. Finally, we also present new, more
accurate estimates of the N\'eel  temperature for both ideal samples and samples containing Mn$_{\rm Cu}$-antisites.
%We checked in this contribution the formation energies of the most probable defects, summarize the influence
%of these defects on the residual resistivity and present also how some defects change the N\'eel temperature.
%We use for all calculations the first-principle electronic structure model, results are compared with available experimental data.

\section{Formalism} \label{Form}

The AFM-CuMnAs prepared by the MBE has a tetragonal structure \cite{cumnas-1,cumnas-str} with the
space group P4/nmm (No.~129) \cite{IntT-A}.
The atomic basis contains two formula units (6 atoms), Cu atoms are in the basal
plane of the tetragonal lattice, (Wyckoff position $2a$), there are two
parallel layers of As atoms (Wyckoff position $2c$) and two layers of Mn atoms
(Wyckoff position $2c$) with oppositely oriented moments.
The experimental lattice parameters are $a$=$b$=3.82~\AA~and $c$=6.318~\AA.
The relative positions of atoms (in units of $c$) are $z_{\rm Cu}$=0.0,
$z_{\rm Mn}$=0.330, and $z_{\rm As}$=0.266.

This geometry is used in the VASP calculations (Vienna ab initio simulation package using the projector
augmented wave scheme \cite{paw}) with the GGA exchange correlation potential \cite{GGA} to evaluate
formation energies of defects.
 The supercell of 96 atoms simulates the sample with defects. For VASP calculations we have used
 $E_{\rm cut}$ = 300 eV, Brillouin zone
sampling with 343 special k-points in the irreducible
three-dimensional wedge of ideal tetragonal CuMnAs structure and correspondingly larger number of k-points in
the supercells of lower symmetry including defects. We used atomic force
minimization to relax the positions of individual atoms in systems with defects.

The transport coefficients and exchange interactions applied in evaluation of resistivity
and the N\'eel temperature are determined using
the Green function formulation of the tight-binding linear muffin-tin orbital (TB-LMTO) method in
which the effect of disorder (defects) is described by the coherent potential approximation (CPA)
\cite{book}. % {\color{blue}
We have verified a good agreement between densities of states of ideal CuMnAs and
alloy with Mn$_{\rm Cu}$-antisites obtained both by the supercell
 VASP and TB-LMTO-CPA calculations \cite{our-PRB} used below
 to estimate the transport properties and exchange interactions.

The transport studies employ the Kubo-Greenwood linear response theory in which the
disorder-induced vertex-corrections are included in the CPA \cite{vertex}.
The effective exchange interactions between Mn atoms for a given shell $s$, $J_{s}$, are
determined by the Lichtenstein mapping procedure \cite{lie} generalized to random alloys
 \cite{iec-our}. Specifically, the disordered local moment (DLM) or paramagnetic reference
 state \cite{our-PRB} was used.

To study the thermodynamic properties of CuMnAs we employed classical Monte Carlo (MC) simulations
based on the Metropolis algorithm \cite{binder}. For simulations we used a 3-dimensional supercell
composed of up to $24 \times 24 \times 24$ elementary CuMnAs cells with periodic boundary conditions.

\section{Results and discussion}

\subsection{Formation energies of defects in AFM-CuMnAs} \label{FE-cumnas}

In previous study \cite{our-PRB} we have determined formation energy of defects assuming a single impurity in
the supercell. This procedure leads to finite supercell magnetic moments for defects including Mn-atoms. In the present
study we have repeated calculations by assuming two Mn-atoms with opposite magnetic moments in the supercell to warrant
its total zero magnetic moment (the supercell size is thus two times larger for the same defect concentration).
The calculated total energy in this model, contrary to previous single impurity one, depends on relative
positions of defects.
We have chosen the most distant position allowed by the supercell size. Specifically, we have employed 96
atoms to simulate 6.25\% defect concentrations.
  While the lattice parameters ($a$, $c$) given by experiment were kept fixed in all cases we have
  optimized atomic positions
inside the supercell (relaxed system).
The accurate determination of formation energies (FE) is a challenging task (see, e.g., a recent review \cite{theory-fe}).
To identify the most probable defects we employ the simple approach in which the FE is defined as
FE=$E_{\rm tot}$[def]
$- E_{\rm tot}$[id] $- \sum_{i} n_{i} E_i$, where $E_{\rm tot}$[def] and $E_{\rm tot}$[id] are
total energies of the supercells with (def) and without (id) defects, $n_{i}$ indicates the number
of atoms of type $i$ ($i$=Cu, Mn, As, vacancy) that have been added to ($n_{i} > 0$) or removed
from ($n_{i} < 0$) the supercell when the defect is formed, and $E_{i}$ are total energies of
atoms in their most probable bulk phase \cite{theory-fe}.
It should be noted that actual values for the FE depend on the choice of these
 energies and on the determination of $E_{\rm tot}$[def].  Results for most common defects are
 given in Table~\ref{t1}.

 %%%%%%%%%%%%%%%%%%%%%%%%%%%%%%%%%%%%%%%%%%%%%%%%%%%%%%%%%%%%%%%%%%%%%%%
\begin{table}[h]
\caption{
The formation energies FE for various substitutional defects in the
tetragonal antiferromagnetic CuMnAs.
The symbol X$_{\rm Y}$ denotes the X-defect on the Y-sublattice.}
\begin{center}
\renewcommand{\arraystretch}{1.2}
\begin{tabular}{cccc}\hline
~~Defect~~ & ~~~FE [eV]~~~
&~~Defect~~ & ~~~FE [eV]~~~
\\\hline
Vac$_{\rm Mn}$& $-$0.16&As$_{\rm Cu}$&+1.73\\
Vac$_{\rm Cu}$&$-$0.14&As$_{\rm Mn}$&+1.79\\
Mn$_{\rm Cu}$&$-$0.03&Mn$_{\rm As}$&+1.92\\
Cu$_{\rm Mn}$&+0.34&Vac$_{\rm As}$&+2.18\\
Cu$_{\rm As}$&+1.15&~~&~~\\%Mn$_{int}$&+1.62\\
\hline
\end{tabular}\renewcommand{\arraystretch}{1.}
\end{center}
\label{t1}
\end{table}
%%%%%%%%%%%%%%%%%%%%%%%%%%%%%%%%%%%%%%%%%%%%%%%%%%%%%%%%%%%%%%%%%%%%%%%

The results show only small shift of FE if the antiferromagnetic elementary cell was used instead of a magnetic one.
  The same qualitative results have been obtained:
 vacancies on Mn- and Cu-sublattices (Vac$_{\rm Mn}$, Vac$_{\rm Cu}$) and Mn$_{\rm Cu}$ and Cu$_{\rm Mn}$ are the most probable defects.
  Mn-interstitials,  Mn$_{\rm As}$ or As$_{\rm Mn}$
 and related defects have large FE. Nevertheless, we note that FE's of defects depend on
delicate details of the impurity kinetics which is not considered here.  We suppose that defects with low
FE's are more probable candidates than those with larger FE's even at the non-equilibrium
conditions.

We note that the quality of samples prepared by MBE is strongly affected by the substrate used
 for growth of the tetragonal CuMnAs film. This complicates the comparison of theoretical results with results
 of X-ray analysis of real samples.
 High concentrations of Cu$_{\rm As}$ and Mn$_{\rm As}$ antisites, defects with larger formation
 energies have been found \cite{cumnas-str} for samples grown on GaAs,
 material with a large mismatch of lattice constants to CuMnAs.

\subsection{Transport properties of AFM-CuMnAs} \label{RHO-cumnas}

In the previous section we have estimated formation energies of
possible defects in the tetragonal CuMnAs.
The primary aim was to indicate the defects responsible for
a finite (planar) resistivity ($\rho_{\rm pl}$ = $\rho_{xx}$ =
$\rho_{yy}$) observed in the experiment.
Here we adopt an alternative strategy and estimate the resistivities
of defects listed in Table~\ref{t1} assuming acceptable defect
concentrations of five percent on each of equivalent sublattices
to determine a suitable defect responsible for a finite sample
resistivity.
We have calculated the planar resistivities for each
defect-type listed in Table~\ref{t1} including also the Cu$\leftrightarrow$Mn swap
defect.
In some cases (see Table~\ref{t3}) we will also present calculated
values of the resistivity normal to the planar one ($\rho_{zz}$) and
the total resistivity
($\rho_{\rm tot}$ = (2~$\rho_{\rm pl}$ + $\rho_{zz}$)/3).
The experimental planar resistivity at very low temperature is about
90~$\mu\Omega$cm.

%%%%%%%%%%%%%%%%%%%%%%%%%%%%%%%%%%%%%%%%%%%%%%%%%%%%%%%%%%%%%%%%%
\begin{table}[h]
\caption{ Calculated planar resistivities $\rho_{\rm pl}$
(in $\mu\Omega$cm) for the tetragonal CuMnAs assuming 5\%
of different defect types on each of equivalent sublattices.
Defect types are the same as those listed in Table~\ref{t1}
with the exception of the Cu$\leftrightarrow$Mn swaps.
}
\begin{center}
\renewcommand{\arraystretch}{1.2 }
\begin{tabular}{ccccc} \hline
{~~~Model~~~} & $\rho_{\rm pl}$ &{~~~Model~~~} & $\rho_{\rm pl}$  \\ \hline
 Vac$_{\rm Cu}$&  12 & As$_{\rm Cu}$&  94 \\
 Vac$_{\rm Mn}$&  36 & As$_{\rm Mn}$&  113 \\
 Mn$_{\rm Cu}$& 111 & Vac$_{\rm As}$& 174 \\
 Cu$_{\rm Mn}$&  24 & Mn$_{\rm As}$&  122 \\
 Cu$_{\rm As}$&  107 & ~~Mn$\leftrightarrow$Cu swap~~& 124 \\
\hline \end{tabular}
\renewcommand{\arraystretch}{1.2 } \end{center} \label{t2} \end{table}
%%%%%%%%%%%%%%%%%%%%%%%%%%%%%%%%%%%%%%%%%%%%%%%%%%%%%%%%%%%%%%%%%

The results are summarized in Table~\ref{t2} with the following
conclusions:
(i) The defects with low formations energies (Vac$_{\rm Mn}$,
Vac$_{\rm Cu}$, Mn$_{\rm Cu}$, and Cu$_{\rm Mn}$) generally give
small planar resistivities as compared to the experiment. The
exception is Mn$_{\rm Cu}$-antisite with a pronounced virtual
bound state at the Fermi energy \cite{our-PRB} which results in larger
resistivity;
(ii) There are relatively large resistivities for defects
on the As-sublattice (Cu$_{\rm As}$, Vac$_{\rm As}$, and
Mn$_{\rm As}$). These large resistivities correspond to
defects with high formation energies; and
(iii) It is interesting to note that the sum of resistivities
of Mn$_{\rm Cu}$ and Cu$_{\rm Mn}$ (135~$\mu\Omega$cm) is
comparable to the resistivity of Cu$\leftrightarrow$Mn swap (124~$\mu\Omega$cm).

In real conditions which are far from the thermodynamical equilibrium
and with possible violation of the sample stoichiometry, the resistivities
depend on the actual occupation of sublattices by the alloy constituents
resulting in the presence of antisite sublattice disorder.
This is a challenging problem for the structural X-ray analysis
in the present alloy because of similar scattering cross sections of atoms
forming the alloy (Cu and Mn).
We have found two such attempts, namely Refs. \cite{cumnas-str} and \cite{carlos}, referred
below as models I and II, respectively.
The first model I \cite{cumnas-str}, simulating a sample grown on GaAs substrate and schematically
written as Cu-(Mn$_{0.86}$,Vac$_{0.14}$)-(As$_{0.84}$,Cu$_{0.08}$,Mn$_{0.08}$),
assumes a small Cu-excess (fully occupied Cu-sublattices with
Cu$_{\rm As}$ antisites), and, at the same time, a reduced Mn-content
with Mn-vacancies on Mn-sublattice (Vac$_{\rm Mn}$) and some Mn$_{\rm As}$ antisites.
While Mn-vacancies have favorable formation energies, the
Cu$_{\rm As}$ antisite and, in particular, Mn$_{\rm As}$ ones have
large formation energies (see Table~\ref{t1}).
We have also tested a model Ia with halved defect concentrations.
The other model II \cite{carlos}, simulating a sample grown on GaP substrate and schematically
written as (Cu$_{1-x}$,Vac$_{x}$)-(Mn$_{1-y}$,Vac$_{y}$)-As has
fully occupied As-sublattice and vacancies on Cu- and Mn-sublattices
with very favorable formation energies.
A detailed analysis \cite{carlos} has lead to two possible structural realizations,
namely model IIa ($x$=11~\% and $y$=4\%) and model IIb ($x$=18\% and
$y$=8\%).
We have also tested a theoretical choice of $x$ and $y$ in the spirit of
the model II, namely the model IIc with $x$=$y$=10\%.
According to \cite{cumnas-2} the samples grown on GaAs are of significantly worse
structural quality as compared to the samples grown on GaP.
The available resistivity data are for the GaP samples.

%%%%%%%%%%%%%%%%%%%%%%%%%%%%%%%%%%%%%%%%%%%%%%%%%%%%%%%%%%%%%%%%%%%
\begin{table}[h]
\caption{ The calculated resistivities $\rho_{\rm pl}$, $\rho_{zz}$ and $\rho_{\rm tot}$
(in $\mu\Omega$cm) for models obtained from the X-ray structural
analysis, namely models I, Ia, IIa, IIb, and IIc (see text for details).
%For a model I
We also show resistivities of separate defects.
% and a model with halved defect concentrations.
%The experimental $\rho_{\rm pl}$ is about
%90~$\mu\Omega$cm measured at $T$=5~K for a sample grown on the
%GaP substrate.
}
\begin{center}
\renewcommand{\arraystretch}{1.2 }
\begin{tabular}{lcccc} \hline
\multicolumn{2}{c}{~~~Model~~~} & $\rho_{\rm pl}$ & $\rho_{zz}$ &
$\rho_{\rm tot}$ \\ \hline
{I} & 14\%Vac$_{\rm Mn}$ 8\% Cu$_{\rm As}$ 8\% Mn$_{\rm As}$&  179 & 213 & 282 \\
{Ia} & 7\%Vac$_{\rm Mn}$ 4\% Cu$_{\rm As}$ 4\% Mn$_{\rm As}$&  148 & 200 & 304 \\
 & 8\% Mn$_{\rm As}$&  141 & 202 & 324 \\
 & 8\% Cu$_{\rm As}$&  129 & 190 & 311 \\
 & 14\%Vac$_{\rm Mn}$&  95 & 138 & 225 \\
\hline
{IIa} & 11\% Vac$_{\rm Cu}$ 4\% Vac$_{\rm Mn}$ &  49  & 171 & 90\\
{IIb} & 18\% Vac$_{\rm Cu}$ 8\% Vac$_{\rm Mn}$ &  89  & 163 & 113 \\
{IIc} & 10\% Vac$_{\rm Cu}$ 10\% Vac$_{\rm Mn}$ &  89 & 182 & 120 \\
\hline \end{tabular}
\label{t3}
\renewcommand{\arraystretch}{1.2 } \end{center} \label{t3} \end{table}
%%%%%%%%%%%%%%%%%%%%%%%%%%%%%%%%%%%%%%%%%%%%%%%%%%%%%%%%%%%%%%%%%%%

Results are summarized in Table~\ref{t3} with the following
conclusions:
(i) In general, resistivities in the direction normal to the plane
($\rho_{zz}$) are larger as compared to the planar ones and so it is
also for the total resistivity $\rho_{\rm tot}$;
(ii) Models of the II-type with Cu- and Mn-vacancies are compatible
with estimated low formation energies (see Table~\ref{t1});
(iii) The experimental determination of exact vacancy concentrations
is not an easy task. The lower planar resistivity of the model IIa in
comparison with the model IIb is related to a very small resistivity
due to Cu-vacancies as compared to that due to Mn-vacancies (3 times
lower, see Table~\ref{t2}).  The explanation is not so easy as different
vacancies also mean differently reduced number of carriers (different
valency of Cu- and Mn-atoms).  Nevertheless, both models IIb and IIc
agree very well with the measured planar resistivity at the very low
temperature (about 90~$\mu\Omega$cm);
(iv) On the other hand, the model I with relatively high concentrations
of Cu$_{\rm As}$ and Mn$_{\rm As}$ antisites with high formation
energies (see Table~\ref{t1}) gives a large planar resistivity, even
with a halved defect concentrations (model Ia).
As it is seen from the Table~\ref{t2}
both defects have separately large planar resistivities. It should be
noted, however, that for the model I grown on the GaAs substrate are no
available transport data; and
(v) For model I we have also compared its resistivity with the
sum of resistivities of separate defects, i.e., with cases of
Cu$_{\rm As}$- and Mn$_{\rm As}$-antisites (each with defect
concentrations of 8\%) and the case of 14\% of Mn-vacancies. Their sum
(365~$\mu\Omega$cm) is two times larger. Such difference has to
be ascribed to different carrier concentrations in the model I and
those in separated cases. In different words, while on one side is the
list of resistivity patterns (Table~\ref{t2}) useful for a qualitative
discussion, it cannot be taken literally for realistic alloys with
complex defect occupations of sublattices with varying carrier
concentrations.

We will now discuss a possible effect of temperature on the
transport.
In addition to the contribution from defects, also phonons (lattice
displacements due to finite temperature) and the spin-disorder (spin
deviations due to temperature) contribute.
A full discussion is beyond the scope of this paper, in addition
such studies are limited to simple lattices \cite{Ebert,Kelly}.
We present results of two simple models, namely the so-called
spin-disorder resistivity (SDR) or the resistivity of the paramagnetic
state above the N\'eel temperature, and the spin-independent
finite-relaxation time (FRT) model which can serve as a qualitative model
for the effect of phonons.
The SDR is simulated by the DLM \cite{dlm,sdr-our} which
is a proper description of the paramagnetic state while the FRT
model is realized by adding a finite imaginary part Im~$z$ to the
Fermi energy in Green functions entering the Kubo-Greenwood formula.
The FRT model has no relations to specific temperature while the
SDR corresponds to the largest possible spin-disorder at the N\'eel
temperature.
The contribution of spin fluctuations well below the N\'eel temperature
is much smaller as compared to the SDR.

The calculated planar SDR's of the ideal CuMnAs sample as well as
of samples with Mn$_{\rm Cu}$-antisites and Cu$\leftrightarrow$Mn swaps (5\%)
are in the range of 225 to 235~$\mu\Omega$cm indicating a weak
influence of chemical disorder.
The SDR is thus larger than the experimental planar resistivity at the
room temperature (300~K, 150~$\mu\Omega$cm).
%{\color{blue}Such result is compatible with much smaller room temperature as
%compared to the N\'eel temperature.}
This could be expected because the N\'eel temperature lies well
above the room temperature.
The spin disorder at room temperature is then rather weak.

We have also calculated resistivity of the model IIc in the framework
of the FRT model for two choices of Im~$z$, namely, Im~$z$=2~mRy and
5~mRy, respectively.
Calculated planar resistivities are 113 and 143~$\mu\Omega$cm,
respectively and they are compatible with the experimental data
at the room temperature assuming some unspecified contribution
due to spin-disorder.

\subsection{Exchange interactions and the N\'eel temperature} \label{EI-cumnas}

The exchange interactions were estimated from the paramagnetic reference
state described by the DLM model.
It describes the finite critical temperature more naturally than the
state at zero temperature (the AFM reference state).
In addition, the paramagnetic state assumes no magnetic order and
the only non-zero interactions are those among Mn-atoms.
We have demonstrated earlier \cite{our-PRB} that for well localized
moments like Mn-ones in CuMnAs, both approaches give similar although
not identical values of exchange integrals.
We have obtained strongly dominating AFM interactions even from
the DLM reference state which are a precursor of the AFM ground state
below the N\'eel temperature.
Calculated exchange interactions were used to construct the effective
classical Heisenberg Hamiltonian from which the N\'eel
temperature was estimated by applying the MC simulations.
Specifically, we have employed the atomistic spin dynamics (ASD)
codes \cite{asd}.
For ideal CuMnAs alloy as well as for systems with vacancies
only Mn atoms on their native sublattices are considered in MC simulations
 while for Mn$_{\rm Cu}$ antisites on Cu sublattice we consider also interactions
between antisite Mn-atoms as well as interactions between antisite and native
Mn-atoms.
The magnetic moments on Mn atoms and the exchange interactions between
Mn atoms in the supercell were set utilizing the results of the
{\it ab initio} calculations.
%%%We defined a 3-dimensional supercell of size $N \times N \times N$,
%%%where in our calculations $N=16$, $20$, and $24$.
The local Mn magnetic moments were assumed to be independent of temperature.
To estimate the average magnetization of the sample and its higher statistical moments
we made use of classical MC method
based on the Metropolis algorithm.

%%%%%%%%%%%%%%%%%%%%%%%%%%%%%%%%%%%%%%%%%%%%%%%%%%%%%%%%%%%%%%%%%%%%%%%%%%%%%%%%%%%%%%%%
\begin{figure}[h]
\begin{center}
\includegraphics[width=70mm,angle=0]{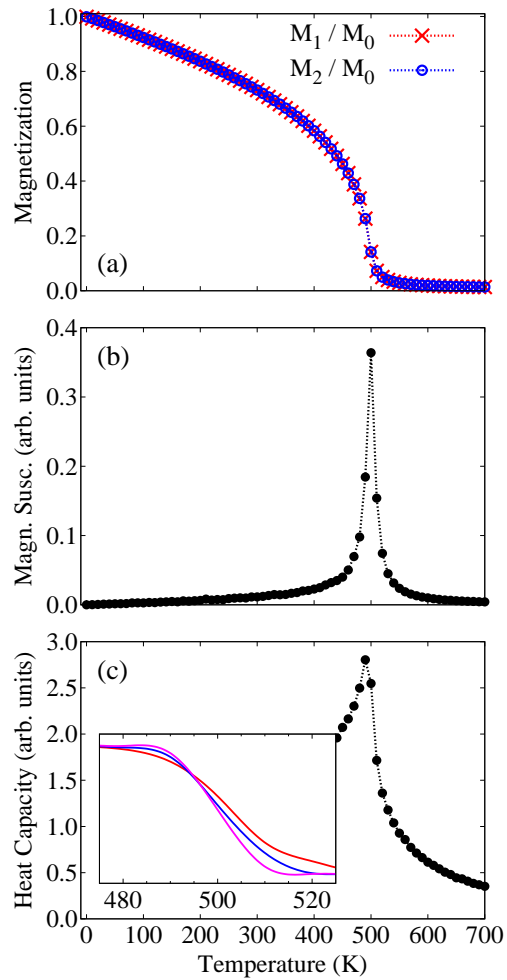}
\end{center}
\caption {(a) Relative magnetizations of the Mn sublattices as a function
of temperature assuming exchange interactions derived from the
paramagnetic (DLM) state of the ideal tetragonal CuMnAs. (b) The magnetic susceptibility
as a function of the temperature
for such CuMnAs alloy.
(c) The temperature dependence
of the heat capacity for this system.  In the inset we show the Binder
cumulants for N = 16, 20, and 24 as a function of the temperature.
 The N\'eel temperature corresponds
to a common intersection of all three curves (495~K).}
\label{f1}
\end{figure}
%%%%%%%%%%%%%%%%%%%%%%%%%%%%%%%%%%%%%%%%%%%%%%%%%%%%%%%%%%%%%%%%%%%%%%%%%%%%%

Results of the MC simulation for ideal tetragonal CuMnAs alloy are shown
in Fig.~\ref{f1}.
In Fig.~\ref{f1}(a) we plot magnetizations of the native Mn sublattices
as a function of the temperature.
As the temperature increases the sublattice magnetizations are reduced due to
spin fluctuations and become zero above the critical N\'eel temperature
$T_{\rm N}$.
The linear decrease of magnetization at low temperatures is a
consequence of the use of the classical statistics. Recently a way
how to include the quantum statistics into MC simulations
has been suggested \cite{berg}.

The N\'eel temperature has been estimated from: (i) The peak in
the magnetic susceptibility [see Fig.~\ref{f1}(b)]; (ii)  the peak in
the heat capacity [see Fig.~\ref{f1}(c)], and (iii) the intersection
of the Binder cumulants \cite{binder1} [see inset of Fig.~\ref{f1}(c)].
The estimate (iii) is considered to be the most accurate estimate of
the N\'eel temperature.
All methods lead to similar results for  the N\'eel temperature, namely
$T_{\rm N}$= 495~K.

%%%%%%%%%%%%%%%%%%%%%%%%%%%%%%%%%%%%%%%%%%%%%%%%%%%%%%%%%%%%%%%%%%%%%%%%%%%%%%%%%%%%%%%%
\begin{figure}[h]
\begin{center}
\includegraphics[width=70mm,angle=0]{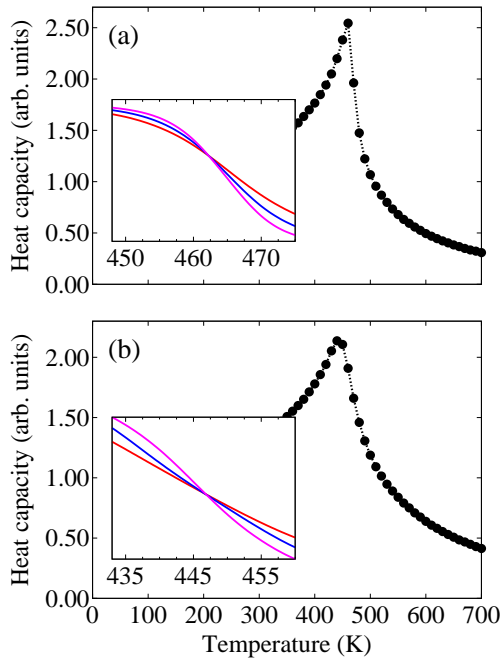}
\end{center}
\caption {The temperature dependence
of the heat capacity derived from the
paramagnetic (DLM) state of the disordered tetragonal CuMnAs alloy: (a)
with Mn$_{\rm Cu}$-antisites (5\%) and (b) with vacancies on Cu and Mn
(model IIa - see text). In the insets are show corresponding Binder
cumulants for N = 16, 20, and 24 as a function of the temperature.
 The N\'eel temperature of 465 K and 446 K has been found.
}
\label{f2}
\end{figure}
%%%%%%%%%%%%%%%%%%%%%%%%%%%%%%%%%%%%%%%%%%%%%%%%%%%%%%%%%%%%%%%%%%%%%%%%%%%%%

We have also studied two alloys in which defects are present, namely
the case of Mn$_{\rm Cu}$ antisites and model IIa containing vacancies
on Mn- and Cu-sites \cite{carlos}.
The latter case is simpler, as we have again only exchange interactions
among Mn-atoms of the native Mn-sublattice, which contains vacancies, i.e.,
Mn atoms are now randomly distributed among Mn lattice sites.
%%The magnetic moments of the native Mn atoms were XXXX????.

 For Mn$_{\rm Cu}$ defects the situation is more complicated.
The supercell is composed of both native Mn atoms organized in two
equivalent fully occupied sublattices and of Mn$_{\rm Cu}$ dopants
distributed randomly on two equivalent Cu-sublattices.
%%In the supercell, positions of the Mn$_{\rm Cu}$ defects were chosen
%%randomly though the Mn atoms are equally divided in both Cu sublattices.
%The magnetic moments of the native Mn atoms were $3.66\, \mu_{\rm B}$,
%and of the Mn$_{\rm Cu}$ antisites $2.45\, \mu_{\rm B}$.

Randomly distributed Mn atoms were modeled by $10$ independent
superlattices with the same amount of Mn atoms. The superlattices
differ by distribution of Mn atoms.
The simulation started from an initial
temperature $900\,{\rm K}$ which decreased by $\Delta{T} = 10\, {\rm K}$. At each
temperature $4\times 10^5$ MC steps were performed.
To improve the statistics, for each of the $10$ configurations of Mn$_{\rm Cu}$ defects
we simultaneously simulated 5 independent identical systems.
The final results were obtained by averaging through all Mn$_{\rm Cu}$ configurations.

The presence of disorder decreased the N\'eel temperature.
The results are shown in Fig.~2 for estimates based on the specific heat
and Binder cumulants, respectively.
 The calculated $T_{\rm N}$ obtained from Binder cumulants are 465~K and
446~K for the case with 5\% of Mn$_{\rm Cu}$-antisites and for the
model IIa, respectively.

\section{Conclusions} \label{Con}

We have performed an extensive ab initio study of electronic, magnetic, and transport
properties of the tetragonal AFM-CuMnAs alloy with potential technological applications.
The VASP approach was used for the estimate of formation energies
of possible defects while the TB-LMTO-CPA method was used to
 calculate transport properties and exchange integrals. Finally,
 the N\'eel temperatures were determined using the Monte-Carlo
 approach from calculated exchange integrals.
The main conclusions are:
(i) The vacancies on Mn- and Cu-
 sublattices, and Mn$_{\rm Cu}$ and Cu$_{\rm Mn}$ antisites %as well as Cu$\leftrightarrow$Mn swaps
  have the lowest formation energies and can be considered as possible candidates for defects
 in CuMnAs; (ii) These predictions are in a good agreement with the X-ray structural analysis
 of samples grown on GaP(001) substrate, in addition, the samples with Cu- and Mn-vacancies have also the
 resistivity close to that found in the experiment; and (iii) We have also obtained
 a good agreement between experimental and calculated N\'eel temperatures. Specifically,
 the vacancies on Mn and Cu as well as the antisite Mn$_{\rm Cu}$ defects reduce the calculated N\'eel temperature
 in comparison with that for the ideal CuMnAs while keeping a good agreement with experiment
 for both quantities.

\section*{Acknowledgements}
We gratefully acknowledge discussions with C. Frontera about the problems of experimental determination
of CuMnAs sample compositions. We acknowledge the financial support from the Czech Science Foundation
(Grant No. 14-37427G). This work was supported by The Ministry of Education, Youth and Sports - project
IT4Inno\-va\-tions National Supercomputing Center – LM2015070 and by the National
Grid Infrastructure MetaCentrum - project LM2015042.


\begin{thebibliography}{99}
\bibitem{cumnas-or} F. M\'aca, J. Ma\v{s}ek, O. Stelmakhovych, X. Marti, K. Uhl\'i\v{r}ov\'a, P.
    Beran, H. Reichlov\'a, P. Wadley, V. Nov\'ak, and T. Jungwirth, Room-temperature
     antiferromagnetism in CuMnAs, J. Magn. Magn. Mater. 324 (2012) 1606.

\bibitem{cumnas-2} P. Wadley, B. Howells, J. \v{Z}elezn\'y, C. Andrews, V. Hills, R.P. Campion,
    V. Nov\'ak, K. Olejnik, F. Maccherozzi, S.S. Dhesi, S.Y. Martin, T. Wagner, J. Wunderlich,
    F. Freimuth, Y. Mokrousov, J. Kune\v{s}, J.S. Chauhan, M.J. Grzybowski, A.W. Rushforth,
    K.W. Edmonds, B.L. Gallagher, and T. Jungwirth, Electrical switching of an antiferromagnet,
    Science { 351}, (2016) 587.

\bibitem{afm-spin} T. Jungwirth, X. Marti, P. Wadley, and J. Wunderlich,
 Antiferromagnetic spintronics, Nat. Nanotech. 11 (2016) 231.

\bibitem{cumnas-1} P. Wadley, V. Nov\'ak, R.P. Campion, C. Rinaldi, X. Mart\'i, H. Reichlov\'a,
    J. \v{Z}elezn\'y, J. Gazquez, M.A. Roldan, M. Varela, D. Khalyavin, S. Langridge, D.
    Kriegner, F. M\'aca, J. Ma\v{s}ek, R. Bertacco, V. Hol\'y, A.W. Rushforth, K.W. Edmonds,
    B.L. Gallagher, C.T. Foxon, J. Wunderlich, and T. Jungwirth, Tetragonal phase of epitaxial
     room-temperature antiferromagnet CuMnAs, Nat. Commun. {4} (2013) 2322.

\bibitem{cumnas-3} L. \v{S}mejkal, T. Jungwirth, and J. Sinova, Route towards Dirac and Weyl
 antiferromagnetic spintronics, Phys. Status Solidi Rapid Res. Lett. {11} (2017) 1700044.

\bibitem{cumnas-str} P. Wadley, A. Crespi, J. G\'azquez, M.A. Rold\'an, P. Garc\'ia, V. Nov\'ak,
    R. Campion, T. Jungwirth,  C. Rinaldi, X. Mart\'i, V. Hol\'y, C. Frontera, and J. Rius,
    Obtaining the structure factors for an epitaxial film using Cu X-ray radiation, J.
    Appl. Cryst. {46} (2013) 1749.

\bibitem{cumnas-tn} V. Hills, P. Wadley, R.P. Campion, V. Nov\'ak, R. Beardsley, K.W. Edmonds,
    B.L. Gallagher, B. Ouladdiaf, and T. Jungwirth, Paramagnetic to antiferromagnetic transition
    in epitaxial tetragonal CuMnAs J. Appl. Phys. { 117} (2015) 172608.

\bibitem{our-PRB} F. M\'aca, J. Kudrnovsk\'y, V. Drchal, K. Carva, P. Bal\'a\v{z}, and I. Turek,
Physical properties of the tetragonal CuMnAs: A first-principles study, Phys. Rev. B 96 (2017)
 094406.

%\bibitem{afm-mnte} M. Krause and F. Bechstedt, Structural and Magnetic Properties of MnTe Phases
%from Ab Initio Calculations, J. Supercond. Nov. Magn. {26} (2013) 1963.

%\bibitem{cu2sb-str} T. Chonan, A. Yamada, and K. Motizuki, Electronic Band Structures of
%Cu2Sb-Type Intermetallic Compounds, J. Phys. Soc. Jpn. {60} (1991) 1638.

\bibitem{IntT-A} International Tables for Crystallography, Volume A: Space-group symmetry,
    edited by Theo Hahn, 5th ed., Kluwer Academic, Dordrecht, Boston, London, 2002.

\bibitem{paw} G. Kresse and D. Joubert, From ultrasoft pseudopotentials to the projector
 augmented-wave method, Phys. Rev. B {59} (1999) 1758.

%\bibitem{VWN} S.H. Vosko, L. Wilk, and M. Nusair, Accurate spin-dependent electron liquid
%correlation energies for local spin density calculations: a critical analysis,
%Can. J. Phys. {58} (1980) 1200.

\bibitem{GGA} J.P. Perdew, K. Burke, and M. Ernzerhof, Generalized Gradient Approximation Made
 Simple, Phys. Rev. Lett. {78} (1997) 1396.

%\bibitem{ldau} S.L. Dudarev, G.A. Botton, S.Y. Savrasov, C.J. Humphreys, and A.P. Sutton,
%    Electron-energy-loss spectra and the structural stability of nickel oxide:  An LSDA+U study,
%    Phys. Rev. B {57} (1998) 1505.

\bibitem{book} I. Turek, V. Drchal, J. Kudrnovsk\'y, M. \v{S}ob, and P. Weinberger,
    Electronic Structure of Disordered Alloys, Surfaces and Interfaces, Kluwer, Boston, 1997.

%\bibitem{bi2te3} K. Carva, J. Kudrnovsk\'y, F. M\'aca, V. Drchal, I. Turek, P. Bal\'a\v{z}, V.
%    Tk\'a\v{c}, V. Hol\'y, V. Sechovsk\'y, and J. Honolka, Electronic and transport properties
%    of the Mn-doped topological insulator Bi$_2$Te$_3$: A first-principles study,
%    Phys. Rev. B {93} (2016) 214409.

\bibitem{vertex} K. Carva, I. Turek, J. Kudrnovsk\'y, and O. Bengone, Disordered magnetic
 multilayers: Electron transport within the coherent potential approximation, Phys. Rev. B {73}
     (2006) 144421.

%\bibitem{rho-our} I. Turek, J. Kudrnovsk\'y, V. Drchal, L. Szunyogh, and P. Weinberger,
% Interatomic electron transport by semiempirical and ab initio tight-binding approaches,
% Phys. Rev. B {65} (2002) 125101.

\bibitem{lie} A.I. Liechtenstein, M.I. Katsnelson, V.P. Antropov, and V.A.  Gubanov,
 Local spin density functional approach to the theory of exchange interactions in ferromagnetic
 metals and alloys, J. Magn. Magn. Mater. {67} (1987) 65.

\bibitem{iec-our} I. Turek, J. Kudrnovsk\'y, V. Drchal, and P. Bruno, Exchange interactions,
 spin waves, and transition temperatures in itinerant magnets, Philos. Mag. {86} (2006) 1713.

\bibitem{binder} K. Binder and D.W. Heermann, Monte Carlo Simulation
 in Statistical Physics, Springer, Berlin, 1997.

%\bibitem{growth} V. Nov\'ak, private communication.

%\bibitem{nimnsb-fe} B. Alling, S. Shallcross, and I.A. Abrikosov, Role of stoichiometric
%and nonstoichiometric defects on the magnetic properties of the half-metallic ferromagnet
%NiMnSb, Phys. Rev. B {73} (2006) 064418.

%\bibitem{cumnsb-fe}  F. M\'aca, J. Kudrnovsk\'y, V. Drchal, I. Turek, O. Stelmakhovych, P. Beran,
%    A. Llobet, and X. Mart\'i, Defect-induced magnetic structure of CuMnSb,
%    Phys. Rev. B {94} (2016) 094407.

\bibitem{theory-fe} C. G. Van de Walle and J. Neugebauer, First-principles calculations
for defects and impurities: Applications to III-nitrides, J. Appl. Phys. {95} (2004) 3851.

\bibitem{carlos} C. Frontera, private communication.

\bibitem{Ebert} H. Ebert, S. Mankovsky, K. Chadova, S. Polesya, J. Min\'ar, and D. K\"odderitzsch,
 Calculating linear-response functions for finite temperatures on the basis of the alloy analogy model,
 Phys. Rev. B 91 (2015) 165132.

\bibitem{Kelly} A.A. Starikov, Y. Liu, Z. Yuan, and P.J. Kelly,
 Calculating the transport properties of magnetic materials from first principles including thermal
 and alloy disorder, noncollinearity, and spin-orbit coupling, Phys. Rev. B 97 (2018) 214415.

%\bibitem{oqmd-fe} S. Kirklin, J.E. Saal, B. Meredig, A. Thompson, J.W. Doak, M. Aykol, S.
%    R\"uhl, and C. Wolverton, The Open Quantum Materials Database (OQMD): assessing the
%    accuracy of DFT formation energies, Comp. Mater. {1} (2015) 15010.

\bibitem{dlm}  B.L. Gyorffy, A.J. Pindor, J. Staunton, G.M. Stocks, and H. Winter,
 A first-principles theory of ferromagnetic phase transitions in metals, J. Phys. F:
    Met. Phys. {15} (1985) 1337.

\bibitem{sdr-our} J. Kudrnovsk\'y, V. Drchal, I. Turek, S. Khmelevskyi,
 J.K. Glasbrenner, and K.D. Belashchenko, Spin-disorder resistivity of ferromagnetic
 metals from first principles: The disordered-local-moment approach,
 Phys. Rev. B {86} (2012) 144423.


\bibitem{asd}  B. Skubic, J. Hellsvik, L. Nordstr\"om, and O. Eriksson, A method for atomistic
spin dynamics simulations: implementation and examples, J. Phys.: Condens. Matter
    {20} (2008) 315203.

\bibitem{berg} L. Bergqvist and A. Bergman, Realistic finite temperature simulations of magnetic
systems using quantum statistics, Phys. Rev. Mater. 2 (2018) 013802.

\bibitem{binder1} K. Binder, Finite size scaling analysis of Ising model block distribution functions, Z. Phys. B 43 (1981) 119.


\end{thebibliography}
\end{document}